# Optical rectification and Pockels effect in the organic nonlinear crystal MBI for quasi-discrete THz time-domain spectroscopy

AGATA M. KOTULSKA,[1] JAKUB MNICH,[1] WEIFENG ZHANG,[2] XINYUAN ZHANG,[2,3] ZHANGGUI HU,[2] NING YE,[2] JAROSŁAW SOTOR,[1] AND ŁUKASZ A. STERCZEWSKI[1,*]

[1]*Laser and Fiber Electronics Group, Faculty of Electronics, Photonics and Microsystems, Wroclaw University of Science and Technology, Wyb. Wyspianskiego 27, 50-370 Wrocław, Poland*

[2]*State Key Laboratory of Crystal Materials, Tianjin Key Laboratory of Functional Crystal Materials, Institute of Functional Crystals, Tianjin University of Technology, Tianjin 300384, China.*

*[lukasz.sterczewski@pwr.edu.pl](mailto:lukasz.sterczewski@pwr.edu.pl)

[3] [xyzhang@email.tjut.edu.cn](mailto:xyzhang@email.tjut.edu.cn)

**Organic nonlinear optical (NLO) crystals offer efficient nonlinear frequency conversion for broadband THz spectroscopy. Unfortunately, only a subset of NLO crystals is compatible with the telecom (~1.5 µm) pump wavelength region, where convenient MHz-rate femtosecond fiber lasers are available. Here, we present ultra-broadband (1–30 THz) THz generation and detection using a benzoindolium iodide (MBI, $C_{24}H_{24}NOI$) crystal illuminated by a 1.5 µm MHz-rate fiber laser. At 200 mW of average pump power, the crystal delivers ~1 µW of THz radiation as quasi-discrete peaks covering the whole emission region with a prominent peak at 11.8 THz. High-dynamic range (>40 dB) time-domain spectroscopy is demonstrated within minutes of integration.**

Although access to the THz range (0.1–10 THz) at room temperature (RT) is possible using thermal sources (globars) or mercury lamps [1], coherent laser frequency conversion schemes including photoconductive switches [2], nonlinear crystals [3], air plasma [4], and spintronic emitters [5] have redefined modern THz spectroscopy in terms of signal-to-noise ratio and temporal resolution capabilities. Recently, much attention has been devoted to organic NLO crystals [6] thanks to their high nonlinear coefficients and compatibility with molecular engineering, which is employed to tune the material's optical properties [7]. This feasibility is critical from a phase-matching perspective, as phase-matching limitations hamper the conversion efficiency and THz generation bandwidth of most known NLO materials. To date, organic NLO crystals have provided broadband THz coverage (1–30 THz) in practical laboratory settings, i.e., without requiring bulky millijoule-class pump lasers. These materials are highly compatible with the most convenient sources of femtosecond pulses: erbium-doped fiber lasers operating at telecommunication wavelengths (~1.5 µm) with MHz rates. Unfortunately, suitable materials for efficient NLO frequency conversion to the THz range from telecom S-, C-, and L-bands remain scarce, and currently include DAST [8], DSTMS [9], HMQ-TMS [10], OH-1 [11], and PNPA [12,13]. Research on new organic NLO materials for THz photonics is now motivated by the sub-optimal emission properties of existing emitters. Excitation of phonons and secondary THz absorption inside the crystal yield a highly modulated, quasi-discrete spectrum. To overcome spectral sparsity, it is possible to coherently combine emission from two spectrally complementary materials, as recently proposed by Rader et al. [14]. This obviously calls for new materials with unique emission features. For instance, spectroscopy of rock minerals often requires coverage at 12 THz (25 µm) [15], where known NLO crystals struggle to emit.

A response to this need – MBI [16] – is a new ionic-type organic NLO crystal, similar to DAST and DSTMS, where the electron-transfer character between the benzindolium-based electron acceptor and electron donor ($-OCH_3$) is responsible for high optical nonlinearities, in particular of the second order $\chi^{(2)}$. Prior experiments have shown the MBI's compatibility with difference frequency generation (DFG) in mJ-energy pulsed excitation using two-color ns pulses tunable between 1.37–1.5 µm [16]. However, the crystal's nonlinear response in the femtosecond regime may not remain the same due to transient molecular dynamics. Here, rather than performing power measurements in the slow (ns) regime, we resolve the ultrafast temporal dynamics of THz pulses generated via optical rectification in the MBI crystal. We leverage the Pockels effect in a THz-induced lensing (TIL) detection scheme to demonstrate the MBI's suitability for ultra-broadband THz time-domain spectroscopy (TDS) [17].

The TDS setup used in our experiment (Fig. 1a) is almost identical to that used in our previous work on PNPA [13]. A $\lambda_c$ = 1.55 µm 18-fs fiber laser with 200 mW of average power and 50 MHz repetition rate is focused onto a 700±10 µm thick MBI crystal to a $1/e^2$ diameter of 75 µm to produce THz pulses resolved in time using the TIL technique, as plotted in Fig. 1b. The orientation of the crystal was (110), effectively employing type-0 phase matching for THz generation and detection. This scheme is very convenient as the pump is incident at normal incidence with polarization identical to the generated (detected) THz. Although the measurement chamber was purged with dry air, residual amounts of water vapor (~2% relative humidity) slightly modulate the emission spectrum below 10 THz (HITRAN model included as a guide to the eye [18]). The THz TDS pulse lasts approximately 2 ps with a noticeable continuous wave component with a period of ~85 fs, corresponding to the dominant quasi-discrete emission peak at ~11.8 THz. To overcome limitations in the detection range of the TDS setup (due to the finite sampling pulse width), we built a Michelson-type THz Fourier transform spectrometer (FTS) (Fig. 1c) with a thin-film lithium tantalate (LTO) detector behind a HDPE window (PR No1, LaserComponents). For uniform THz waveform sampling (and hence proper frequency calibration), both setups utilized an additional mechanically coupled, auxiliary interferometer measuring fringes from a $\lambda = 1.54932$ µm continuous wave (CW) reference laser when the scanning mirror was moved. Fig. 1d plots the THz interferogram and its Fourier transform measured by the FTS instrument. Despite the lower spectral dynamic range (due to incoherent, non-gated detection), the Fourier spectrum of the FTS interferogram shows identical peaks to THz-TDS spectrum. However, the spectral roll-off above 15 THz is not as severe as in the TDS because convolution losses from the sampling pulse do not concern FTS. Since the THz-TDS spectrum represents a convolution between the THz emitter and detector response, THz absorption in the MBI crystal serving as a detector deepens notches in the TDS spectrum and enhances emission peaks in regions where it can emit. This explains why the spectra agree in peak position and envelope, but not in the relative depth of the emission peaks.

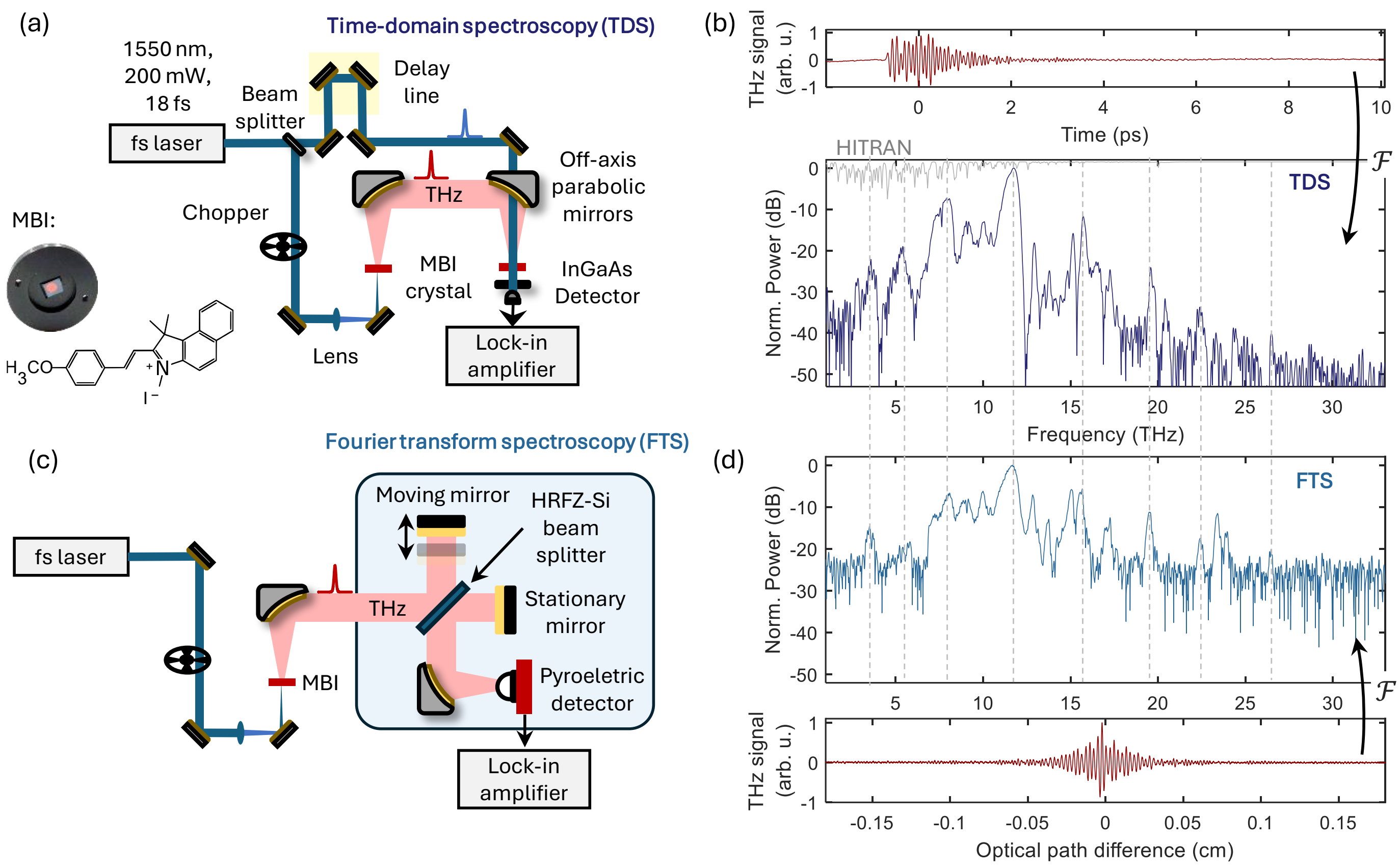

**Fig. 1.** (a) Experimental setup of the THz time-domain spectrometer. (b) THz field-resolved time-domain signal and its Fourier spectrum. Emission from the MBI crystal is probed by another one via THz-induced lensing. The measured spectrum was averaged 50 times for a lock-in time constant of 1 ms and a delay line scan speed of 1 mm/s. (c) Setup of the Fourier transform spectrometer probing the emission spectrum from the MBI pumped at 1.55 µm. (d) THz emission spectrum of the MBI crystal measured via FTS under dry-air conditions. The spectrum was averaged 10 times for a lock-in time constant of 30 ms and a delay line scan speed of 10 µm/s. Using two complementary measurement setups allows confirmation that spectral features are real (non-artifact). In each setup, a suitable pump-blocking filter was used with residual pump transmission of $10^{-7}$.

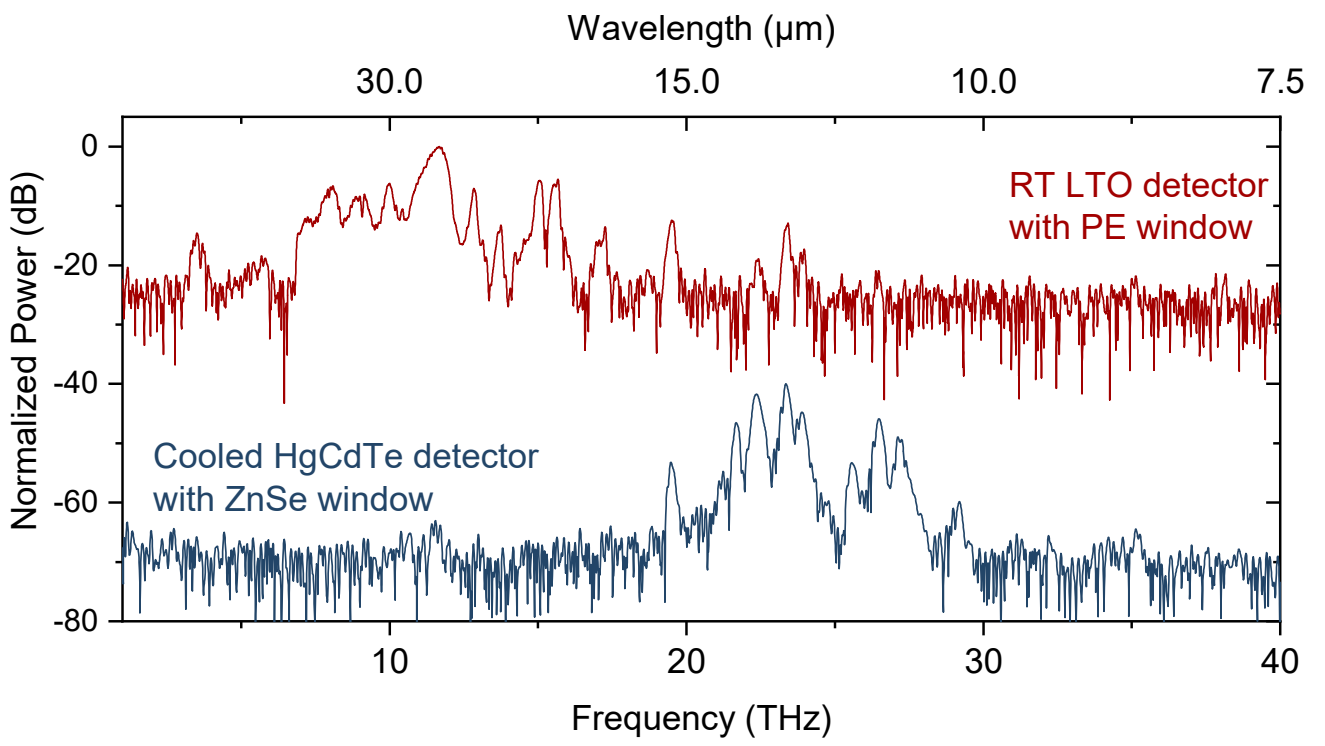

**Fig. 2.** Fourier spectra measured using a RT lithium tantalate (LTO) pyroelectric detector with a polyethylene (PE) window and a highly refractive zone silicon beam splitter (top curve, same as Fig. 1d for visual comparison of peak location), along with a vertically offset spectrum when a KRS-5 beam splitter was used in tandem with a thermoelectrically-cooled HgCdTe detector.

The FTS setup features the ease of changing the detector and beam splitter; therefore, by replacing the previously used highly refractive float-zone beam splitter and LTO detector with a bulk thallium bromoiodide (KRS-5) substrate (operating from the visible range down to 50 μm of wavelength), and changing the detector to a longwave-infrared HgCdTe (Vigo, PVI-4TE-10.6), we detected peaks between 20–30 THz (Fig. 2) with improved sensitivity. Although the two detection schemes are optimized for different spectral ranges (THz vs longwave infrared – LWIR), they yield spectra that agree very well in regions where their detection characteristics overlap. Weak emission peaks in the LTO-detected spectrum above 20 THz become very prominent in the spectrum probed by the HgCdTe detector (~30 dB of signal-to-noise ratio). The fact that those peaks are detectable using the FTS instrument with two *different* photodetectors, and remain visible in the TDS spectrum, confirms that the nominally weak high-frequency spectral features are not just noise floor artifacts of the TIL technique but rather real emission features due to proper group-phase index matching and lack of strong absorption at these frequencies (wavelengths). Therefore, the MBI crystal is expected to be used in the community to generate quasi-discrete LWIR and THz radiation at the at the frequencies summarized in Table 1.

**Table 1. Identified MBI emission peaks in THz units from TDS and FTS measurements for pumping at $\lambda_c$=1.55 μm using sub-20 fs pulses. Bolded values in brackets were obtained via a difference generation (DFG) process using ns lasers (ref. [16]).**

| | | | |
|---|---|---|---|
| 3.5 | 5.4 | 8.0 | 9.1 |
| 10.0 | 11.8 (**12.3**) | 12.9 (**12.3**) | 13.8 |
| 15.1 | 15.7 (**16.1**) | 17.3 | 19.5 |
| 20.4 | 21.7 | 22.4 | 23.4 |
| 23.9 | 25.6 | 26.5 | 27.2 |
| 29.2 | 35.2 | | |

*Conversion to wavelength in μm is given by: λ (μm) ≈ 300 / ν (THz)*

To better understand the discrete spectral nature of the emission peaks from the MBI crystal, we studied its linear optical properties in the near-infrared and THz regions governing the type-0 phase matching employed in our experiment. First, we measured the near-infrared group index using the FTS instrument. We inserted the 650±10 μm-thick crystal into the stationary arm of the interferometer with all long-pass filters removed. This arm imbalance delayed and chirped the interferogram enabling group index assessment. For optimal near-infrared performance, we replaced the HgCdTe detector with an amplified InGaAs detector while keeping the KRS-5 beam splitter in the beam path. For THz absorption and refractive index measurements, we used the TDS system with a pair of DSTMS crystals (also using the TIL technique). Strong absorption of the MBI at the highest THz frequencies restricted our analysis to the 2–12 THz range.

Figure 3 plots the characterization results with the MBI crystal oriented as (110). The fs laser emission range (1.4–1.7 μm, Fig. 3a) converts to a quasi-discrete THz spectrum with a series of peaks with gradually increasing intensities up to ~11.8 THz (Fig. 3b). The degraded efficiency of low-frequency THz generation and detection relates to diffraction losses (shown with a dashed line) reaching ~20 dB at 2 THz. This is because in this region, the near-infrared spot size is smaller than that of the generated THz field. These losses also result from the hole in parabolic mirrors, as discussed in Ref. [19].

We then merged the measured pump-wavelength MBI's group index of ~2.33 with minor deviations (Fig. 3c) with the complex refractive index (absorption coefficient and refractive index) in the THz range plotted in Fig. 3d to understand the emission spectrum. As a guide to the eye, we include 3 characteristic regions denoted by Roman numerals (I, II, III). In region I, phase matching $\Delta n = n_g - n_{THz}$ is almost perfect (shaded blue region), yet losses from diffraction (~7 dB) and strong secondary absorption render a weak emission peak (–19 dB relative to maximum). In region II (7–8 THz), $\Delta n$ becomes pronounced (0.03), yet low absorption and reduced diffraction losses (~3 dB) yield a much stronger and flatter peak (–7 dB relative to max.). Finally, the strongest peak in region III at ~12 THz results from optimizing all factors: a relatively low absorption coefficient, excellent phase matching, and negligible diffraction losses (0.5 dB). Notably, this analysis shows that a low absorption coefficient is critical for efficient THz generation and detection. In the sub-mm crystal-thickness regime, one can say that it is even more important than perfect phase matching. For instance, between 3–4 THz $\Delta n$ reaches an unfavorable 0.07, but because of the low absorption, the emission peak at 3.5 THz is of comparable strength to the one at 5.4 THz in regime (I), where perfect phase matching is accompanied by 3× higher absorption. Unfortunately, additional losses from diffraction at 3–4 THz (6 dB stronger than at 5 THz) attenuate the low-frequency peak.

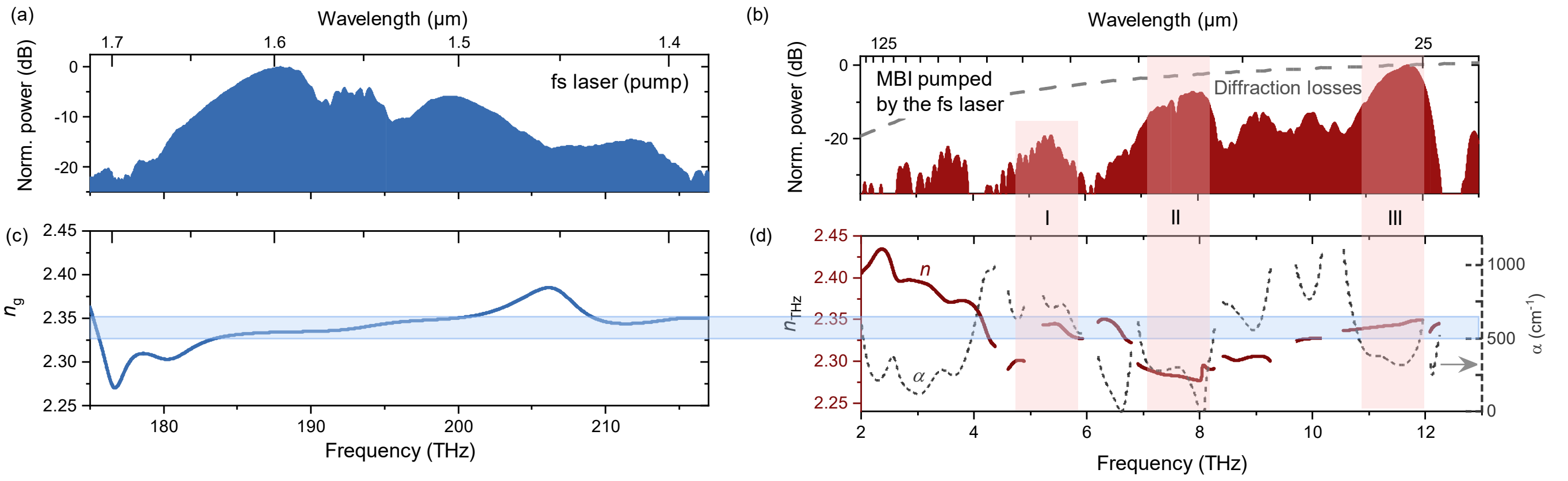

**Fig. 3.** (a) Spectrum of the fs laser measured using the FTS setup. (b) THz spectrum measured using TDS plotted for phase-matching analysis (identical to that in Fig. 1b). (c) Group index of MBI measured using the FTS setup. (d) Refractive index (*n*) and absorption coefficient (*α*) of the MBI crystal measured using a DSTMS-based TDS spectrometer (which is not gap-less and has regions of no intensity – hence the gaps in the dispersion and absorption spectra. The panels depict the spectral structure of the THz emission spectrum of MBI. The group index at the pump wavelength is approximately $n_g \approx 2.33$. When the THz refractive index $n_{THz}$ approaches this value, phase matching is obtained. However, the second effect is absorption of generated THz radiation in the crystal. Panels (b) and (d) show three distinct regimes: (I) when = $n_g - n_{THz}$ is low, but α is high – despite phase matching, absorption is strong, and only a weak peak is obtained. (II) $\Delta n$ is large, but absorption is low – a medium-intensity peak is observed. (III) Nearly perfect phase matching $\Delta n \approx 0$ with relatively low absorption α – the strongest emission peak is generated. The third effect is losses from diffracting THz radiation due to a small pump (probe) spot size (dashed line in panel (b)). These lossess are strongest at low THz frequencies.

To quantify the power conversion efficiency $\eta$ of MBI, we measured the average THz power after a long-pass optical filter (Knight Optical, 5000FSL2505-C) followed by a 20 THz mesh filter (Swiss Terahertz). We use the LTO detector as the power sensor connected to a lock-in amplifier. We obtained ~1 μW of average power, corresponding to $\eta_M = 5 \times 10^{-6}$. Using the same set of filters, we also measured ~1 μW from a 700 μm-thick DSTMS crystal (Swiss Terahertz). This suggests that MBI matches the nonlinear frequency-conversion performance of state-of-the-art nonlinear frequency converters, yet its highly discrete spectrum enhances the power spectral density in selected spectral regions.

In summary, we studied optical rectification and the linear Pockels effect in benzoindolium iodide. This organic nonlinear crystal enables high-dynamic-range (>40 dB) THz field-resolved spectroscopy with ultra-broadband spectral coverage reaching 30 THz when pumped by MHz-rate nJ pulses at telecom wavelengths. Analysis of the group and refractive indices highlights the importance of a low absorption coefficient, which should be prioritized in the computationally assisted search of new nonlinear materials for broadband THz generation and detection [12].

**Funding.** European Research Council (101117433, ERC Starting Grant, TeraERC). The National Natural Science Foundation of China (Grant No. 52272007).

**Acknowledgment.** A. Kotulska, J. Mnich, and L. A. Sterczewski acknowledge funding from the European Union (ERC Starting Grant, TeraERC, 101117433). X. Zhang acknowledges funding from The National Natural Science Foundation of China (Grant No. 52272007). Views and opinions expressed are, however, those of the authors only and do not necessarily reflect those of the European Union or the European Research Council Executive Agency. Neither the European Union nor the granting authority can be held responsible for them. This work is supported by the use of National Laboratory for Photonics and Quantum Technologies (NLPQT) infrastructure, which is financed by the European Funds under the Smart Growth Operational Programme.

**Disclosures.** The authors declare no conflict of interest.

**Data availability.** The data that support the findings of this study are publicly available in Ref.[20].

## REFERENCES

1. E. K. Plyler, D. J. C. Yates, and H. A. Gebbie, J. Opt. Soc. Am. **52**, 859 (1962).
2. D. H. Auston, K. P. Cheung, and P. R. Smith, Applied Physics Letters **45**, 284 (1984).
3. X. -C. Zhang, X. F. Ma, Y. Jin, T. -M. Lu, E. P. Boden, P. D. Phelps, K. R. Stewart, and C. P. Yakymyshyn, Appl. Phys. Lett. **61**, 3080 (1992).
4. X. Xie, J. Dai, and X.-C. Zhang, Phys. Rev. Lett. **96**, 075005 (2006).
5. T. Seifert, S. Jaiswal, U. Martens, J. Hannegan, L. Braun, P. Maldonado, F. Freimuth, A. Kronenberg, J. Henrizi, I. Radu, E. Beaurepaire, Y. Mokrousov, P. M. Oppeneer, M. Jourdan, G. Jakob, D. Turchinovich, L. M. Hayden, M. Wolf, M. Münzenberg, M. Kläui, and T. Kampfrath, Nature Photon **10**, 483 (2016).
6. M. Jazbinsek, U. Puc, A. Abina, and A. Zidansek, Applied Sciences **9**, 882 (2019).
7. D. Kim, W. T. Kim, J.-H. Han, J.-A. Lee, S.-H. Lee, B. J. Kang, M. Jazbinsek, W. Yoon, H. Yun, D. Kim, S. van Bezouw, J. Campo, W. Wenseleers, F. Rotermund, and O.-P. Kwon, Advanced Optical Materials **8**, 1902099 (2020).
8. A. Schneider, M. Stillhart, and P. Günter, Opt. Express, OE **14**, 5376 (2006).
9. M. Stillhart, A. Schneider, and P. Günter, J. Opt. Soc. Am. B, JOSAB **25**, 1914 (2008).
10. F. D. J. Brunner, S.-H. Lee, O.-P. Kwon, and T. Feurer, Opt. Mater. Express **4**, 1586 (2014).
11. M. Jazbinsek, U. Puc, A. Abina, and A. Zidansek, Applied Sciences **9**, 882 (2019).
12. C. Rader, Z. B. Zaccardi, S.-H. E. Ho, K. G. Harrell, P. K. Petersen, M. F. Nielson, H. Stephan, N. K. Green, D. J. H. Ludlow, M. J. Lutz, S. J. Smith, D. J. Michaelis, and J. A. Johnson, ACS Photonics **9**, 3720 (2022).
13. L. A. Sterczewski, J. Mnich, and J. Sotor, Advanced Physics Research **4**, 2400105 (2025).
14. C. Rader, M. F. Nielson, B. E. Knighton, Z. B. Zaccardi, D. J. Michaelis, and J. A. Johnson, Opt. Lett. **47**, 5985 (2022).
15. J. Ciazela, J. Bakala, M. Kowalinski, B. Pieterek, M. Steslicki, M. Ciazela, G. Paslawski, N. Zalewska, L. Sterczewski, Z. Szaforz, M. Jozefowicz, D. Marciniak, M. Fitt, A. Sniadkowski, M. Rataj, and T. Mrozek, Frontiers in Earth Science **11**, (2023).
16. W. Zhang, Z. Wang, X. Zhang, Y. Wang, D. Xu, Z. Hu, J. Yao, and Y. Wu, Crystal Growth & Design **22**, 3311 (2022).
17. A. Schneider, I. Biaggio, and P. Günter, Appl. Phys. Lett. **84**, 2229 (2004).
18. I. E. Gordon, L. S. Rothman, R. J. Hargreaves, F. M. Gomez, et al., Journal of Quantitative Spectroscopy and Radiative Transfer **353**, 109807 (2026).
19. U. Puc, T. Bach, P. Günter, M. Zgonik, and M. Jazbinsek, Advanced Photonics Research **2**, 2000098 (2021).
20. A. Kotulska and L. A. Sterczewski, "Dataset for “Optical rectification and Pockels effect in the organic nonlinear crystal MBI for quasi-discrete THz time-domain spectroscopy”, Figshare (2026)," https://doi.org/10.6084/m9.figshare.35015608.